# Automated detection of oral pre-cancerous tongue lesions using deep learning for early diagnosis of oral cavity cancer.

Mohammed Shamim[*,1], Sadatullah Syed[2], Mohammad Shiblee[3], Mohammed Usman[1] and Syed Ali[4]

[1]Dept. of Electrical Engineering, College of Engineering, Abha, King Khalid University, Saudi Arabia

[2]Dept. of Diagnostic Sciences and Oral Biology, College of Dentistry, King Khalid University, Saudi Arabia

[4]Deanship of University Development, Taif University, Saudi Arabia

[3]Computer Engineering Department, King Khalid University, Saudi Arabia

*Email: mzmohammad@kku.edu.sa

**ABSTRACT**

Discovering oral cavity cancer (OCC) at an early stage is an effective way to increase patient survival rate. However, current initial screening process is done manually and is expensive for the average individual, especially in developing countries worldwide. This problem is further compounded due to the lack of specialists in such areas. Automating the initial screening process using artificial intelligence (AI) to detect pre-cancerous lesions can prove to be an effective and inexpensive technique that would allow patients to be triaged accordingly to receive appropriate clinical management. In this study, we have applied and evaluated the efficacy of six deep convolutional neural network (DCNN) models using transfer learning, for identifying pre-cancerous tongue lesions directly using a small dataset of clinically annotated photographic images to diagnose early signs of OCC. DCNN model based on Vgg19 architecture was able to differentiate between benign and pre-cancerous tongue lesions with a mean classification accuracy of 0.98, sensitivity 0.89 and specificity 0.97. Additionally, the ResNet50 DCNN model was able to distinguish between five types of tongue lesions i.e. hairy tongue, fissured tongue, geographic tongue, strawberry tongue and oral hairy leukoplakia with a mean classification accuracy of 0.97. Preliminary results using an (AI + Physician) ensemble model demonstrate that an automated initial screening process of tongue lesions using

DCNNs can achieve "near-human" level classification performance for detecting early signs of OCC in patients.

## 1. INTRODUCTION

OCC is one of the most common oral malignancies and accounts for almost 3% of all cancer cases diagnosed worldwide [1]. According to the World Health Organization, more than 650,000 cases of OCC are reported each year [2]. OCC is very much prevalent in individuals mostly from developing countries due to lack of awareness and limited access to clinical diagnosis and dental specialists, predominantly in South Central Asia [2]. OCC normally manifests on the lips, gums, tongue and inner lining of the cheeks, roof and floor of the mouth [3]. A patient suffering from severe OCC has difficulties with eating, speaking, appearance of lumps in the oral cavity, physical marks on the face due to surgical procedures and treatment, severe pain etc [4]. If OCC is allowed to progress undetected it can easily spread to other parts of the body such as neck and lungs. More than 80% of the OCC's are preceded by manifestation of lesions that are referred to as oral potentially malignant lesions (OPMD) [5]. OPMDs are generally considered as initial signs of many systemic disorders and a number of oral diseases. Early diagnosis of these OPMD would reduce their chances of cancerous transformation [6]. Here we have focused only on lesions occurring on the tongue. There are several types of tongue lesions, hence identifying which OPMD has tendency to transform into oral cancer is challenging, and requires specialized training. This problem is further compounded by lack of sufficiently trained physicians and dental specialist in developing countries especially in remote areas, where OCC is widespread. An efficacious and automated classification method of pre-cancerous tongue lesions can both support physicians in their daily clinical duties and individuals with limited medical expertise with fast and inexpensive access to life saving diagnosis via apps on mobile platforms [7].

Recent successes have ushered a new age of computational intelligence for automated diagnosis in healthcare. Due to their inherently agnostic nature, deep learning techniques are successfully transforming medical image analysis in pathology, radiology and other fields of medical imaging. Esteva et al and Hosny et al, have both utilised DCNNs for classifying cancerous skin conditions with reported accuracy's of 72.1% and 98.61% respectively [8,9]. Lakhani and Sundaram utilized an ensemble of DCNNs to classify pulmonary tuberculosis with an accuracy of 99%. [10]. An extensive search of the various research databases (e.g. Google Scholar, ScienceDirect etc) was performed for any research work with sufficient scientific proceeding related to application of deep learning for detection of OCC. Here we present a few for brevity. L. Ma et al, presented an OCC classification model using DCNNs to detect head and neck cancer in mice using hyperspectral images with an average accuracy of 91.36% [11], H. Rajaguru utilized Gaussian Mixture Measures and Multi-Layer Perceptron to classify recurrence of oral cancer using features extracted manually from medical databases and achieved a maximum accuracy of 94.18% [12]. D. W. Kim et al, demonstrated a deep learning based survival prediction model of oral cancer patients using medical record database with an accuracy of 78.1% [13]. K. Lalithamani, used data mining techniques to develop a deep neural based adaptive fuzzy system OCC classification model with an accuracy of 96.29% [14] H. Wieslander et al, utilized convolutional neural networks (CNNs) to classify OCC cell images obtained using PAP-smear tests with prediction accuracy in the rage of 78–82% [15]. J. Folmbsee explored active training methods to efficiently train CNNs on pathological OCC tissue images with a predication rate greater than 93% [16]. P. R. Jayaraj reported an accuracy of 91.4% by utilizing hyperspectral images of cancerous tumours to classify their malignancies using CNNs [17]. M. Aubreville et al, classified OCC cells using laserendomicroscopy images of the oral cavity using CNNs with a mean accuracy of 88.3% [18]. All of the reported studies either utilize confidential patient medical records database to extract demographic, clinical, and histopathological features or utilize cell/pathological images obtained using specialised clinical procedures and advanced imaging systems that are neither cost-effective nor readily available in developing countries. Hence there is an urgent need of a generalised DCNN based classification model capable of accurately classifying oral malignancies directly from readily available photographic images. To the best of our knowledge there is no published work directly utilizing clinically annotated photographic images of tongue lesions for classification study. The main contributions of this paper are as follows,

1) An AI based computational method that can automatically detect oral pre-cancerous tongue lesions directly from clinically annotated photographic images, to aid physicians/dentists in their early diagnosis before their manifestation into cancerous malignancies.

2) Pre-trained DCNNs based on Vgg19 and ResNet50 architectures classified tongue lesions with high binary (98 %) and multiclass (97 %) classification accuracies respectively.

3) Using a physician augmented approach; we achieved "near-human" level classification performance for detecting early signs of OCC in patients.

## 2. BACKGROUND

The tongue is one of the most important structures of the oral cavity that controls critical functions of eating, tasting, and speaking. Assessment of the tongue has historically been a critical part of any clinical examination as several congenital and pathological lesions occur exclusively on the tongue. A basic and thorough knowledge of commonly occurring tongue lesions can aid the medical practitioner/dentist to diagnose the severity of the lesion and effectively manage the patient.

Oral lesions occurring on the tongue are enormous and can present a diagnostic and therapeutic dilemma for physicians. They can be differentiated based on their clinical characteristics such as size, location, surface morphology, colour etc [19]. Majority of the oral lesions are short-lived that resolve with simple medical treatment (benign), while some can cause long term difficulties (pre-cancerous / cancerous). In such cases, taking a complete history and a thorough oral examination by a physician is highly recommended. Here, we have summarised a few of the most commonly occurring superficial tongue lesions for brevity in Fig. 1. Oral Thrush (OT), Fissured Tongue (FT), Geographic Tongue (GT), Hairy Tongue (HT), Pigmented Fungiform Papillae (PFP) and Strawberry Tongue (ST) are clinically annotated as benign lesions, which resolve with correction of the underlying condition and do not pose a health risk to the individual if treated at an early stage [20]. Oral Hairy Leukoplakia and Erythroplakia are clinically annotated as pre-cancerous with high risk of cancerous transformation [20]. Traditionally lesions classed as pre-cancerous or cancerous need to be identified at a very early stage in order to prevent extensive local invasion of the oral cavity. Such lesions normally require treatment in the form of radiation therapy and even surgery [20].

| Tongue Lesions (Training Class) | Clinical Cause | Clinical Features | Clinical Classification |
|---|---|---|---|
| 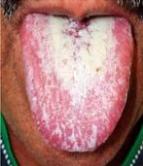 Oral thrush (OT) | Often associated with fungus *Candida albicans infection* on the tongue. | Thick, white or creamy coloured deposits (spots). | Benign |
| 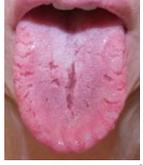 Fissured Tongue (FT) | Associated with Melkersson's Rosenthal syndrome and Down syndrome. | Cracks of varying depth and sizes appear on the top and edges of the tongue. | Benign |
| 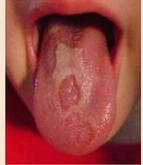 Geographic Tongue (GT) | Cause is unknown. | Red area of varying sizes surrounded by irregular white border. | Benign |
| 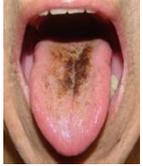 Hairy Tongue (HT) | Extensive use of antibiotics, radiation treatment, bacteria growth etc. | Hair-like appearance of varying colour on the top of the tongue. | Benign |
| 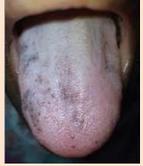 Pigmented Fungiform Papillae (PFP) | Cause is unknown. | Presence of dark spots on the tongue. Common amongst people with dark skin. | Benign |
| 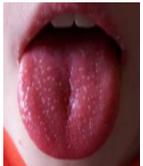 Strawberry Tongue (ST) | Normally associated with Kawasaki disease, Scarlet Fever or Vitamin B-12 deficiency. | Tongue appears swollen, red with bumps, like a strawberry. | Benign |
| 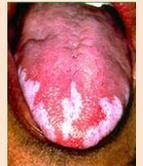 Leukoplakia (LP) | Associated with human immunodeficiency virus infection (HIV) | Whitish area or spots mostly on the lateral border of the tongue that cannot be scraped off. | Pre-cancerous |
| 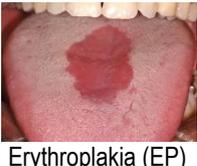 Erythroplakia (EP) | Associated with heavy smoking, tobacco chewing, excessive alcohol use etc. | Raised or smooth fiery red patch that often bleeds when scraped. | Pre-cancerous |

**Fig. 1:** Summary of a few common superficial tongue lesions [20].

DCNNs are one of the most popular artificial intelligence algorithms used for deep learning. A DCNN model learns to perform classification task directly from images, videos, texts and even sound. DCNNs are particularly useful for finding patterns in images for object detection and classification studies. They are capable of learning directly from image data, using patterns to classify images, thereby eliminating the need for manual feature extraction. Such an approach has proven to produce state-of-the-art recognition results in object recognition tasks and computer vision applications such as document recognition [21] etc.

There are several popular DCNN architectures that are available for different applications. Table 1, summarises a few of the popular DCNN models utilized in this study. These models were primarily selected due to their popularity in the world of medical image classification and also to cover a range of models with varying number of layers, training parameters and input image sizes. Majority of these DCNN models are trained on a subset of the ImageNet database [22] that was used for the ImageNet Large-Scale Visual Recognition Challenge [23]. These pre-trained DCNNs have learned to extract powerful and informative features from images in 1000 categories of the ImageNet database. As shown in Table 2, DCNNs have different number of layers (depth), i.e. the total number of sequential convolutional layers from the input layer to the final output classification layer and different number of parameters. The inputs to all the models are RGB images of varying width ($W$), height ($H$) and a fixed depth (D) of 3 for red, green and blue channels. In this study, we will evaluate the efficacy of transfer learning of these pre-trained DCNN models for detecting pre-cancerous oral tongue lesions.

| Year | DCNN Models | No. of Layers (Depth) | Input Image Size ($W \times H \times D$) | No. of Parameters (Millions) | Reference |
|------|-------------|------------------------|------------------------------------------|-------------------------------|-----------|
| 2012 | AlexNet | 8 | 227×227×3 | 61.0 | [24] Krizhevsky et al |
| 2014 | GoogLeNet | 22 | 224×224×3 | 7.0 | [25] Szegedy et al |
| 2014 | Vgg19 | 19 | 224×224×3 | 144 | [26] Simonyan et al |
| 2015 | Inceptionv3 | 48 | 299×299×3 | 23.9 | [27] Szegedy et al |
| 2015 | ResNet50 | 50 | 224×224×3 | 25.6 | [28] He et al |
| 2016 | SqueezeNet | 18 | 227×227×3 | 1.24 | [29] Iandola et al |

**Table 1**: Summary of pre-trained DCNNs used in this study.

## 3. MATERIALS & METHOD

*3.1     Tongue Lesion Dataset*

To the best of our knowledge, there is no available dataset of lesions occurring in the oral cavity. Hence a custom dataset was created using clinically annotated photographic images. The images of the different oral tongue lesions were collected from the Internet using an image search engine. The images are of varying sizes, lighting condition and taken from different perspective of the medical practitioner, making the images inherently augmented in nature. The lesion images and training classes illustrated in Fig. 1 were manually annotated by a certified physician with more than 15 years of clinical practice before being cleared for model training. All of the images were then converted into Joint Photographic Experts Group (JPEG) format as required by our deep neural framework. They were then resized to the required input image size of the DCNN models (Table 1) prior to the model training process. To improve model accuracy by training them on the region of interest (ROI) i.e. tongue, additional and unwanted areas of the images, facial features such as nose, eyes etc were cropped. This process also de-identified the images for patient confidentiality.

*3.2     Transfer learning of pre-trained DCNNs*

Traditionally, a very large number of training images are needed to train a DCNN from scratch. To overcome the problem of relatively small number of training images available in our tongue lesion dataset, transfer learning of DCNN is utilized. Transfer learning allows for utilising DCNN models pre-trained on a large dataset to new problems with limited data. In essence, we take the last few layers from a pre-trained DCNN model and "fine-tune" them on a new dataset (Fig. 2). Fine tuning a pre-trained network is often faster and much easier than constructing and training a neural network from scratch, as it allows the network to learn features specific to the new dataset. In this study, transfer learning of pre-trained AlexNet, GoogLeNet, Vgg19, ResNet50, Inceptionv3 and SqueezeNet models is achieved by modifying the final classification layer class parameters for binary ($N=2$) and multiclass ($N=5$) inference tests. The pre-trained convolutional (feature extraction) layers along with their trained weights are transferred directly to the new proposed models. To learn faster in the new classification layers we set the weight and bias learning rate factors to 20. A small global learning rate of 0.0001 is used for the transferred layers so as to not change their weights dramatically. The stochastic gradient descent (SGD) algorithm with momentum of 0.9 was utilised for updating the weights of the network. All the DCNN models are trained using back propagation.

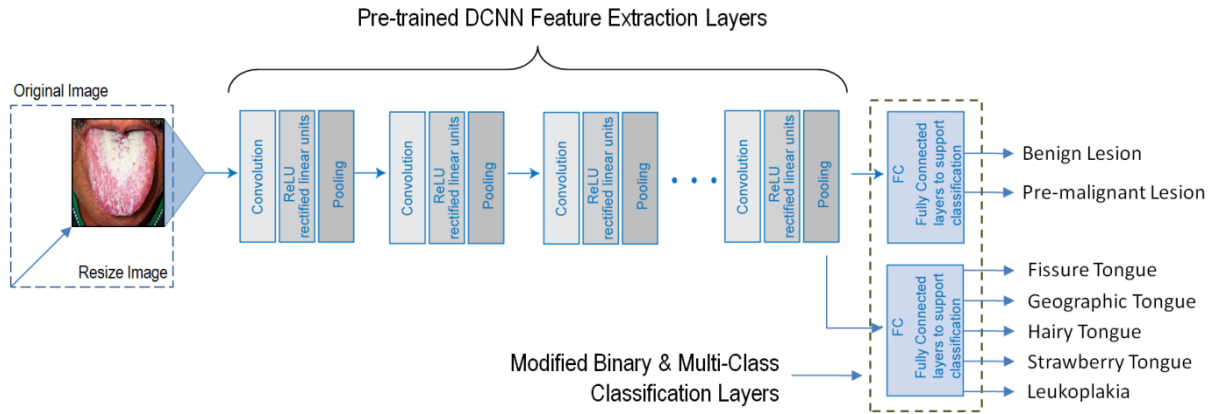

**Fig. 2:** Mechanism of transfer learning of DCNNs and the applicable scope of the prediction classes.

*3.3      Experimentation*

All of the experiments were performed using a desktop computer equipped with a six core Core i5 (8$^{th}$ Gen.) processor with 16 GB of DDR4 RAM and a NVIDIA GeForce GTX-1060 GPU (6 GB GDDR5 graphics memory and 1280 CUDA cores). Matlab 2018b (64-bit) along with Deep Learning Toolbox (DLT) and Parallel Computing Toolbox (PCT) were used for coding the tongue lesion classification experiments. DLT was utilized for transfer learning of pre-trained DCNN models. PCT allows for training the new DCNN model using the GPU. Model training on a GTX-1060 GPU is significantly faster than on a Core i5 CPU due its parallel architecture that utilizes 1280 CUDA cores for parallel computing. We validate the effectiveness of the DCNN models in two ways. First, using binary *(N=2)* classification of tongue lesions as benign or pre-cancerous and second using multiple class *(N=5)* classification (4 benign and 1 pre-cancerous). These inferences test the DCNN models ability to detect pre-cancerous tongue lesions. Additionally, in cases where the highest scoring algorithm misclassifies an image, an ensemble approach (AI + Physician) is utilized where the misclassified image is further verified by a certified physician.

Due to relatively small number of images in our dataset, for binary classification inference test (*N=2*), 200 images were used. This dataset composed strictly of tongue lesions that are annotated as either benign or pre-cancerous. For the multi-class inference test *(N=5),* a total of 300 images from four benign classes i.e. HT, FT, GT, ST and one pre-cancerous class i.e. LP were used. An equal number of images from each of the training classes were utilized to create balanced datasets for both the inference tests. In both the tests, images were randomly split into training (80%) and validation images (20%). The training set was utilized for generating prediction models whereas the validation images were used to estimate the model accuracy's. Both the training and validation images were shuffled every epoch before each network validation. Online image augmentation is performed by arbitrarily, random flipping of the training images along the vertical and horizontal axis with 50% probability during training. This technique ensures that the network sees a completely different dataset every epoch, thereby preventing the DCNNs from overfitting and memorizing the exact details in the tongue lesion training images. However, the augmented images are not used for validation, as it is not possible to completely eliminate the possibility of the lesion images having some inherent properties that are rotation-variant.

For a qualitative and quantitative analysis of all the DCNN models trained on the tongue lesion dataset, we computed the mean model prediction Accuracy ($A_{CC}$), Specificity ($S_{PEC}$) the true negative rate and Sensitivity ($S_{ENS}$) the true positive rate [30] from the generated confusion matrix after as listed in equations 1 - 3. Additionally model training time ($T_{SEC}$) was also monitored.

$$A_{CC} = \frac{t_p + t_n}{t_p + t_n + f_p + f_n} \quad (1)$$

$$S_{ENS} = \frac{t_p}{t_p + f_n} \quad (2)$$

$$S_{PEC} = \frac{t_n}{t_n + f_p} \quad (3)$$

Here *tn*, *tp*, *fp* and *fn* refer to true negative, true positive, false positive and false negative. All the DCNN model training parameters (batch size, number of epochs) were tuned to achieve the highest $A_{CC}$. Misclassification by the highest scoring model is further evaluated by a 2<sup>nd</sup> certified physician with more than 15 years of clinical experience for target class verification. For each misclassified image, the physician was tasked to classify them as either benign or pre-cancerous *(N=2)* or their respective target class *(N=5)*. The physician outputs a single prediction per image.

## 4. RESULTS & DISCUSSION

*4.1 Predictive performance of different DCNN models for binary classification of pre-cancerous tongue lesions.*

For classifying tongue lesions as benign or pre-cancerous *(N=2)*, prediction models based on AlexNet, GoogLeNet, Vgg19, Inceptionv3, ResNet50 and SqueezeNet architectures were built using 160 training images (80%) and their performance was evaluated on 40 validation images (20%). The six models were evaluated by computing the mean values of their performance metrics (i.e. $A_{CC}$, $S_{ENS}$, and $S_{PEC}$) from the generated confusion matrix (Table 2). The generated confusion matrix allows for visual evaluation of the DCNN model in correctly classifying validation images into their respective target classes. Further, the measured DCNN model sensitivity's and specificity's were plotted using the receiver operating curve (ROC) space with 50% discrimination threshold and the area under the curve (AUC) was computed to further assess model performance. Additionally, the ensemble approach was utilized for the misclassified images of the highest scoring model to evaluate them as either benign or pre-cancerous and its corresponding confusion matrix was also generated for comparative analysis.

| Model | Binary Classification *(N=2)* | | | |
|---|---|---|---|---|
| | $A_{CC}$ (%) | $S_{ENS}$ (%) | $S_{PEC}$ (%) | $T_{SEC}$ (sec) |
| AlexNet | 0.93 ± 0.06 | 0.88 | 0.94 | 50.30 |
| GoogLeNet | 0.93 ± 0.02 | 0.80 | 0.88 | 83.27 |
| ResNet50 | 0.90 ± 0.04 | 0.84 | 0.96 | 188.42 |
| Vgg19 | 0.98 ± 0.04 | 0.89 | 0.97 | 212.09 |
| Inceptionv3 | 0.93 ± 0.03 | 0.83 | 0.88 | 372.95 |
| SqueezeNet | 0.93 ± 0.09 | 0.85 | 0.96 | 48.84 |

**Table 2:** Predictive *(N=2)* performance of DCNN models. Each field shows the mean values; in addition the accuracy field is represented with ± standard deviation over multiple training executions of the DCNN models and their respective $T_{SEC}$.

In our binary classification test, Vgg19 architecture consistently performed the best among the six DCNN prediction models, with highest mean $A_{CC}$ of 0.98 ± 0.04 in 212.09 secs (Table 2). Meanwhile AlexNet, GoogLeNet, Inceptionv3 and SqueezeNet DCNN models all reported a mean $A_{CC}$ of 0.93 in 50.30 secs, 83.27 secs, 372.95 secs and 48.84 secs respectively. ResNet50 was the only model that achieved the lowest mean $A_{CC}$ of 0.90 ± 0.04 in 188.42 secs. It is worth mentioning, that the SqueezeNet model recorded the fastest $T_{SEC}$ when compared to similar performing DCNN models (Table 2). This is exceptionally fast and efficient, considering that the SqueezeNet model has only 1.24 million training parameters, which is significantly lower than all the DCNN models utilized in this study (Table 1).

In Fig. 3, we compare the confusion of matrix of the highest scoring Vgg19 model and the ensemble (AI + Physician) approach. Of the 40 validation images, Vgg19 model correctly classified all the 20 benign lesions; however the model misclassified 1 out of 30 pre-cancerous lesion images as benign (Fig. 3a). Using the ensemble model, the misclassified image was then blindly verified by the physician as being a pre-cancerous lesion (Fig. 3b). Hence by using an ensemble approach we are able to achieve 100 % binary $A_{CC}$ as shown in Fig. 4. Using the ROC curve we plot the mean $S_{ENS}$ and (1–$S_{PEC}$) of the all DCNN models and compute the AUC as shown in Fig. 5. The computed AUC's of the ROC curves of the DCNN models were based on the validation dataset, which had not been seen by the trained DCNN models. The ROC curve of the DCNN models is represented in red whereas the shaded green and red areas graphically represent the AUC of each of the DCNN models. Here, Vgg19 model achieved the highest AUC of 0.9896 followed by GoogLeNet, SqueezeNet and Inceptionv3 all achieving an AUC of 0.9714 (Fig 5). AlexNet and ResNet50 both achieved the lowest AUC of 0.9677 and 0.9529 respectively.

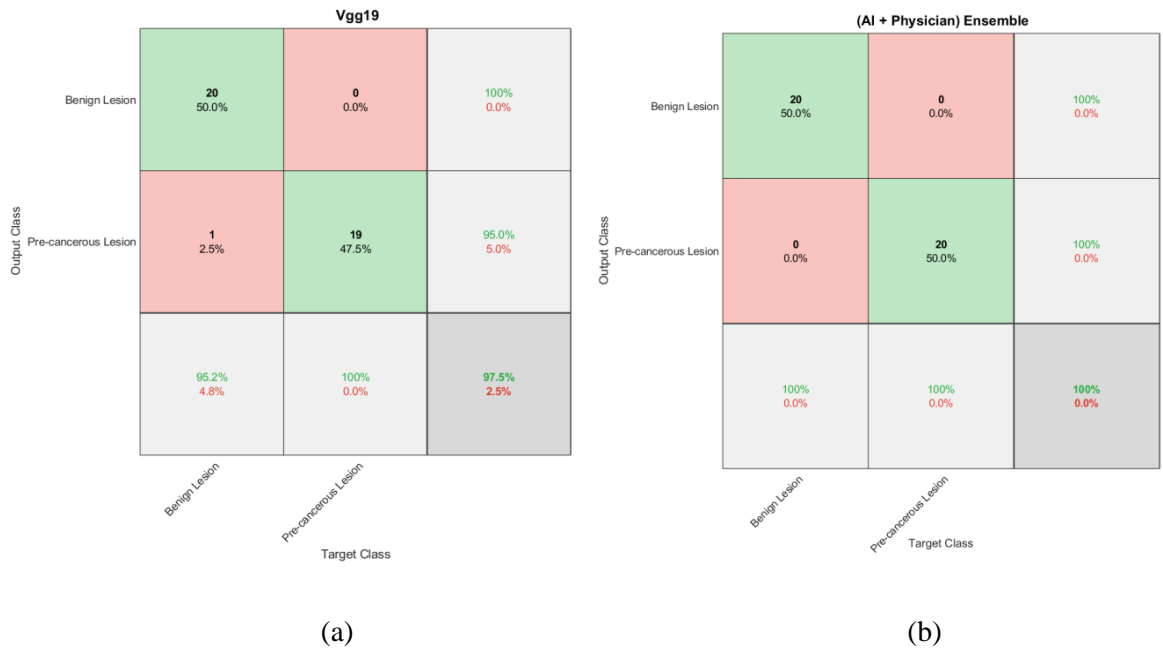

(a)                          (b)

**Fig. 3:** Confusion matrix for a) trained Vgg19 model and b) ensemble (AI + Physician) model for classifying tongue lesions as benign or pre-cancerous. Red boxes represent every misclassified image class and the green box represents each correctly classified class. The output class on the y-axis is the prediction performance of the DCNN model against each target class on the x-axis. Ideally all correctly classified images should appear on the diagonal axis, as is the case with the ensemble model.

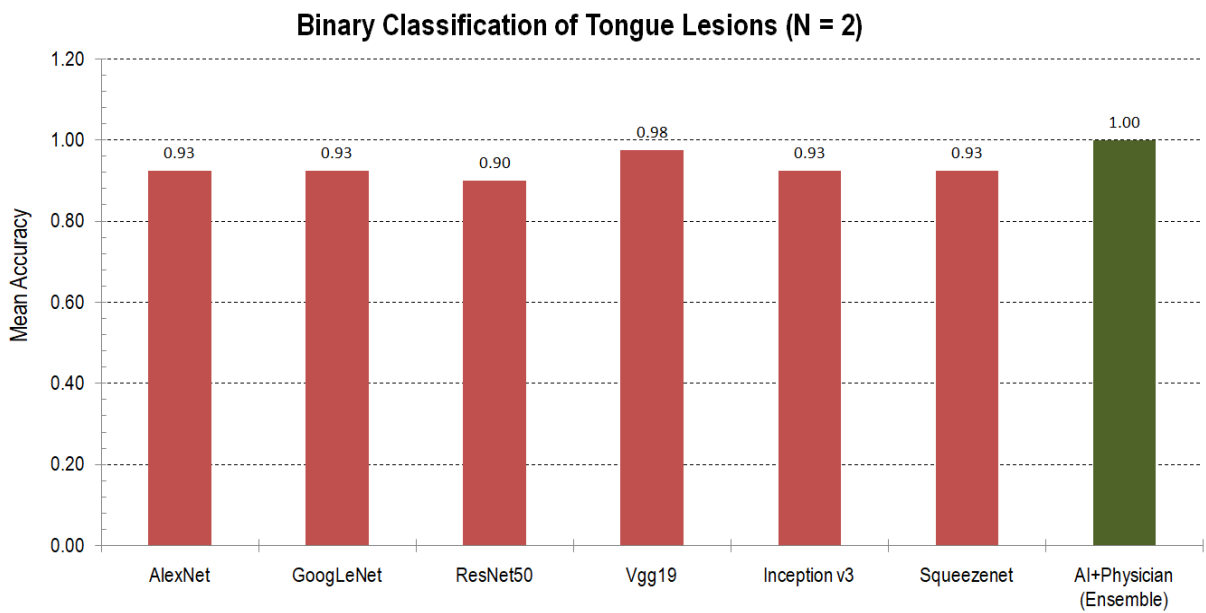

**Fig. 4:** Mean *(N=2)* $A_{CC}$ of the six DCNN models and ensemble model for detecting pre-cancerous tongue lesions.

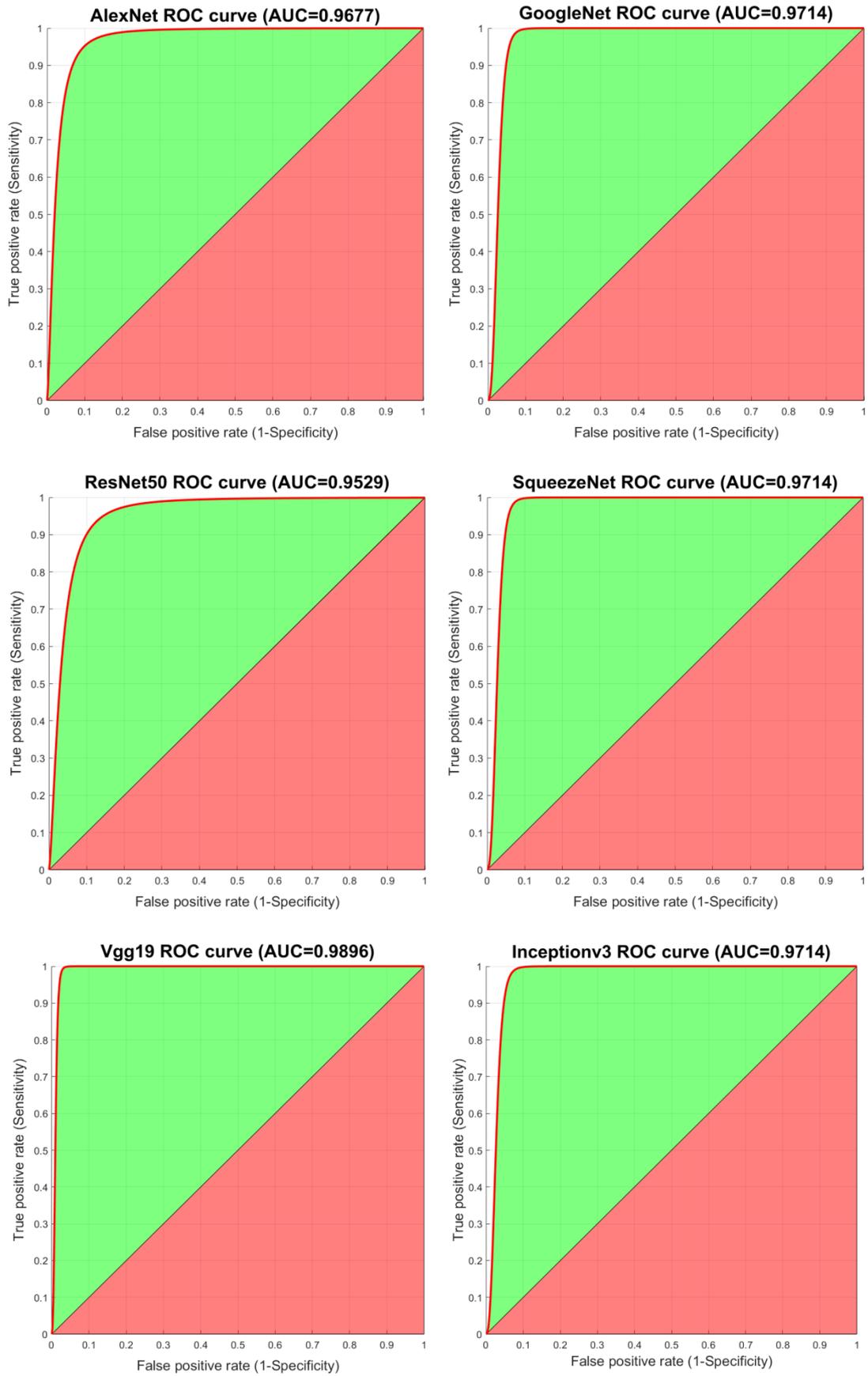

**Fig. 5:** ROC plots and computed AUC of the trained DCNN models on validation dataset.

*4.2 Predictive performance of different DCNN models for 5-class classification of tongue lesions.*

To detect pre-cancerous tongue lesions, multiclass prediction models *(N=5)* based on AlexNet, GoogLeNet, Vgg19, Inceptionv3, ResNet50 and SqueezeNet architectures were built using 240 training images (80%) and their performance was evaluated on 60 validation images (20%). The six models were evaluated by computing the mean $A_{CC}$ and generating a five target class confusion matrix. As shown in Table 3, the ResNet50 model was observed to consistently perform the best among the six prediction models with highest mean $A_{CC}$ of (0.97 ± 0.03) in 225.80 secs followed by Vgg19, Inceptionv3 and SqueezeNet with mean $A_{CC}$ of (0.95 ± 0.03) in 256.88 secs, (0.92 ± 0.02) in 447.27 secs and (0.90 ± 0.04) in 57.58 secs respectively. Lastly, AlexNet and GoogLeNet both reported lowest mean $A_{CC}$ of (0.83 ± 0.08) in 60.64 secs and (0.88 ± 0.06) in 97.58 secs respectively (Table 3).

| Model | Multi-Class Classification *(N=5)* | |
|---|---|---|
| | $A_{CC}$ (%) | $T_{SEC}$ (sec) |
| AlexNet | 0.83 ± 0.08 | 60.64 |
| GoogLeNet | 0.88 ± 0.06 | 97.58 |
| ResNet50 | 0.97 ± 0.03 | 225.80 |
| Vgg19 | 0.95 ± 0.03 | 256.88 |
| Inceptionv3 | 0.92 ± 0.02 | 447.27 |
| SqueezeNet | 0.90 ± 0.04 | 57.58 |

**Table 3:** Predictive *(N=5)* performance comparison of DCNN models. Each field shows the mean $A_{CC}$ with ± standard deviation of the DCNN over multiple training executions along with their respective $T_{SEC}$.

The mean $A_{CC}$ achieved here, were relatively low when compared to those achieved in the binary classification study (Table 2), this can be attributed to relatively small number of training images (48 images per training class) used for multiclass classification study. The reduction in $A_{CC}$ can also be attributed to that fact that not all augmentation strategies are guaranteed to improve $A_{CC}$ all cases and require further investigation to study this effect. As previous, the ensemble approach (AI + Physician) was utilized for the misclassified images of the highest scoring ResNet50 model and further evaluated by a certified physician into their correct target class and the corresponding confusion matrix was also generated for comparative analysis (Fig. 6). The generated confusion matrix allows for visual evaluation of the DCNN model in correctly classifying each of the 60 validation images into their respective target class. Each element (x,y) of the confusion matrix represents the empirical probability of the output class *y*, given that the target class is *x*. Using the multiclass *(N=5)* image dataset, the ResNet50 model achieved the highest mean $A_{CC}$ of 0.97 ± 0.03 while misclassifying two GT lesions as ST lesions (Fig. 6). Using the ensemble approach we then further evaluated the two misclassified images into their correct target classes, to achieve a maximum $A_{CC}$ of 100 % (Fig. 7). Classification errors by the ResNet50, Inceptionv3 and Vgg19 models can be considered less critical as the misclassified lesions were all benign in nature. However, we can see that AlexNet, GoogLeNet and SqueezeNet all of them misclassified pre-cancerous lesions as being benign which can be dangerous and also life threatening for the patient. Hence to overcome this ambiguity, the ensemble approach can be utilized for successfully pre-screening patients to be triaged accordingly to receive appropriate clinical care.

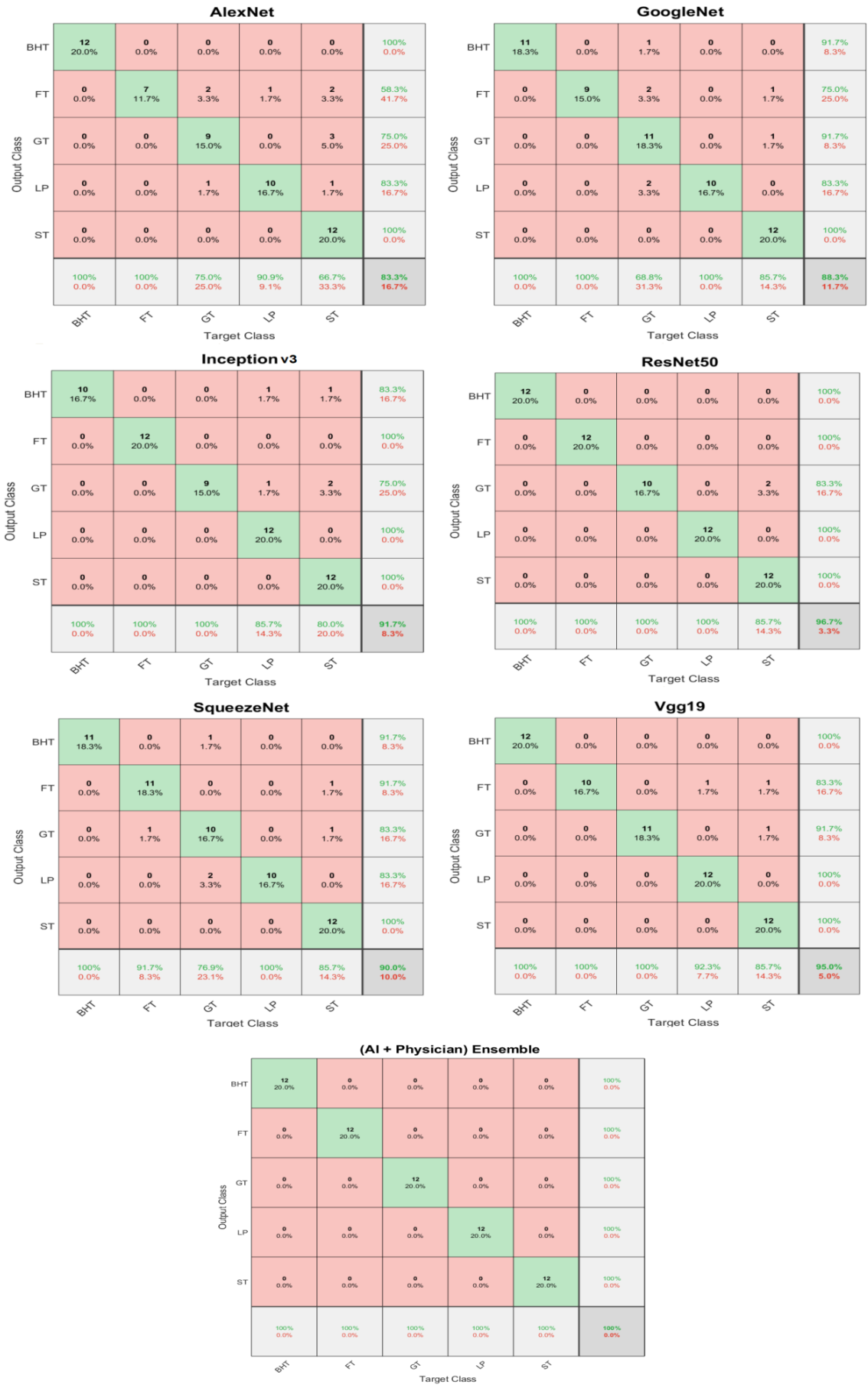

**Fig. 6:** Confusion matrix of the six trained DCNN models and the ensemble model for classifying 60 unseen validation images into 5 target classes.

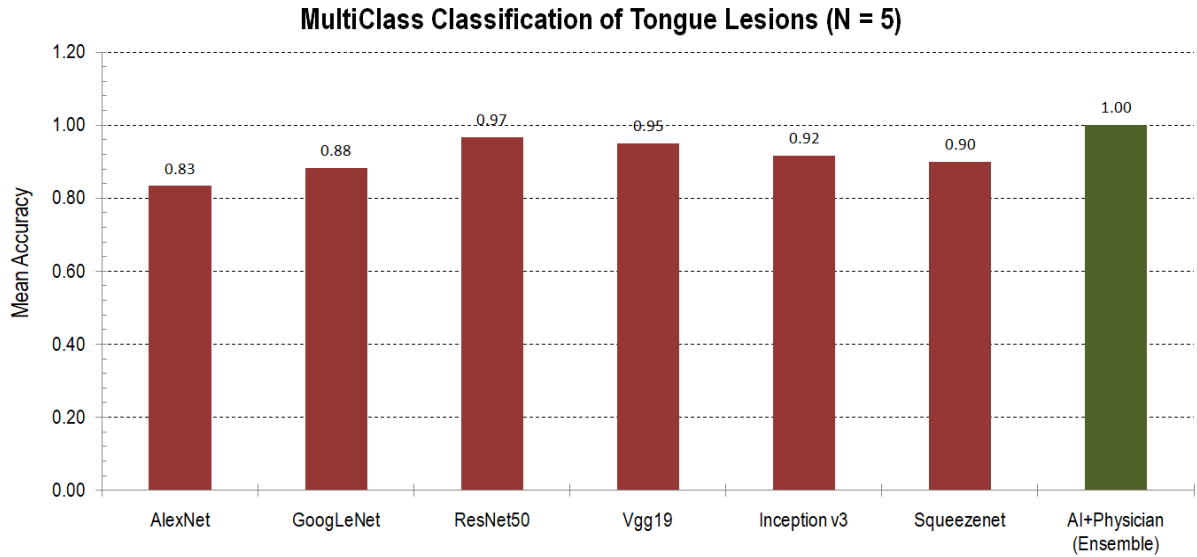

**Fig. 7:** Mean *(N=5)* $A_{CC}$ of the six DCNN models and ensemble model for detecting pre-cancerous tongue lesions.

These tests highlight the difficulty of automated detection discernment of this medically critical tongue lesions classification problem. Given, the limited training dataset, we have demonstrated here that it is possible to achieve "near-human" level classification performance (Figs. 4 and 7) of pre-cancerous tongue lesions using an ensemble approach. Deep learning is a sub-branch of artificial intelligence in which computers perform automated classification task by directly analysing the relationship within the images by employing multiple hidden layers [24-29]. In this study we have utilized supervised DCNNs, which has been extremely successful in several reported medical image classifications [8-11, 15, 16]. A key advantage of using DCNN models is their ability to excel with high dimensional data, such as images, that can be represented at several levels of complexity i.e. from pixel intensity, to edges, to parts of an image and finally the object as a whole. In this study, pre-trained DCNN models trained on just clinically annotated photographic images of tongue lesions achieved near-human level performance at detecting OPMDs that have the tendency to transform into OCC (Figs. 4 and 7). The results achieved here, in some cases, exceed those published earlier using clinical images generated by expensive imaging equipments and clinical procedures that are not necessarily easily available in developing countries [15-18].

Transfer learning of DCNN models pre-trained on everyday images is a viable technique to aid in classification tasks of medical images. Although, the use of DCNN models trained on non-medical images to aid classification of medical images at first may not seem intuitive. There are however similarities, as medical images share similar elements such as edges and pixel intensities that compose the initial feature extraction layers in a network. By transferring learned parameters from pre-trained networks, fully connected layers were initially set to random weights initialization in order to relearn from the tongue lesion images. Previous works [31] have suggested that by supplying more image variants to the DCNN models can improve generalization and performance of the DCNN models. It would be interesting to evaluate the efficacy of additional training images and other augmentation strategies on the $A_{CC}$ of tongue lesions which will be explored in future studies.

A major issue with deep learning models is over-fitting [32]. This normally occurs when the trained DCNN model is not able to generalise well to unseen images, but fits well to the original trained data. This is more apparent with small sets of training images, as in our case. All the DCNN models used in this study have utilized dropout and regularization layers to avoid this problem [24,25,32]. From Fig. 8, it is apparent that training loss and validation loss are similar which indicates well-fit curves. In case of over-fitting, the training loss would have been much higher than the validation loss. In addition, the inherent augmented nature of the original images plus the utilization of online augmentation techniques aided the DCNN models to generalise well and should provide reasonably accurate results on unseen tongue lesions images.

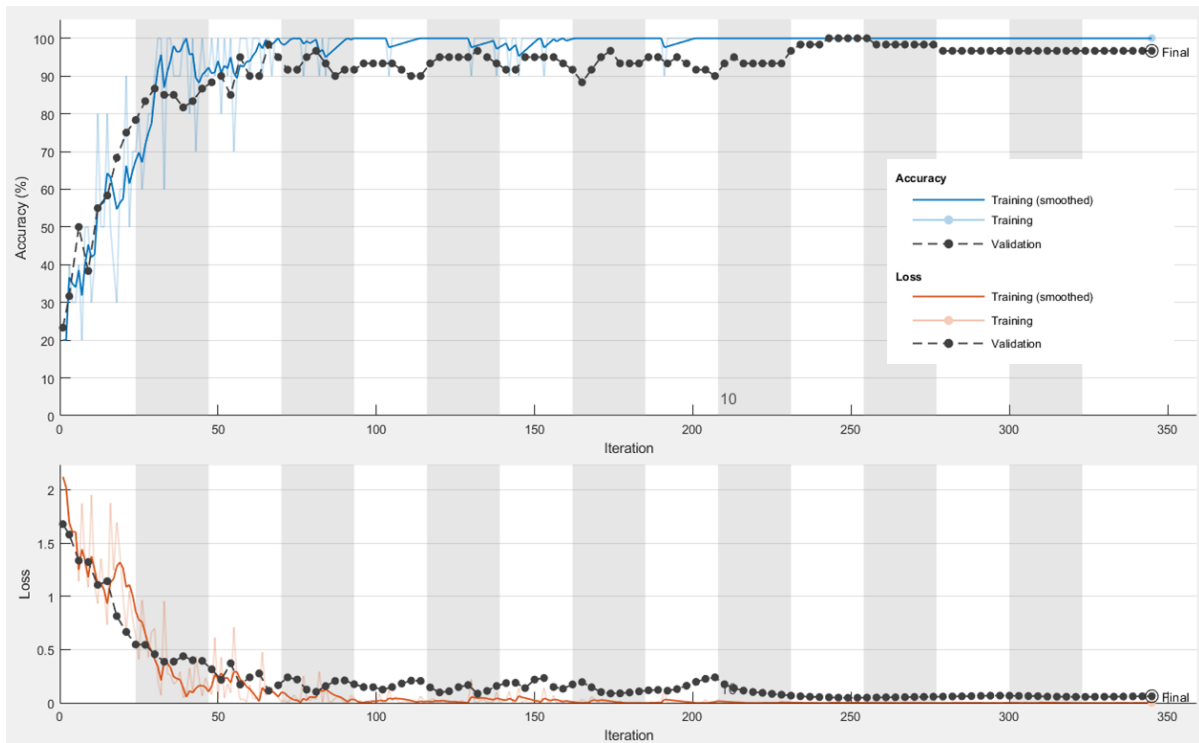

**Fig. 8:** Training curve of ResNet50 DCNN classifier for a 5-class classification problem. The top half of the graph represents the accuracy over the course of the training that increases over time, with a final validation $A_{CC}$ of ~ 97 % (black curve) at the final epoch. The two blue curves in the top accuracy chart represent the classification and smoothed classification accuracy on each individual mini-batch. Training was performed for 15 epochs (23 iterations per epoch) where each epoch represents one pass through the entire training dataset. The two orange curves in the bottom loss chart represent the classification and smoothed classification loss on each individual mini-batch. The bottom black curve represents the loss on the validation set. These training curves are used for selecting the best DCNN model suited for our tongue lesion dataset.

It should be noted that DCNNs in this study outperformed those published in earlier works [14–18], despite having only a small number of training images for both the *(N=2)* and *(N=5)* classification studies. It is not clear if this is due to the different DCNN architectures, fine-tuning strategies or augmentation strategies used in this work and needs further investigation. Also, as with multiple medical studies that utilise DCNNs [23, 28, 33-36], the tongue lesion images were down-sampled to lower resolutions before being fed to the neural networks. Using higher resolution images may improve $A_{CC}$, especially for detecting subtle differences in lesions, but will dramatically increase the $T_{SEC}$ and will require more robust computing platforms and graphical processing units.

DCNNs are normally described as functional black boxes [37], as it is very difficult to determine as to how the network arrived at its conclusion. This is a crucial consideration as one would want to know if the DCNN was actually looking at the lesion on the tongue or rather at other non-relevant parts of the image per se. The complexity of the functional black box is compounded by the sheer size of the trained parameters (e.g. 144 million in the case of the Vgg19 architecture), and it would be practically infeasible to analyse each of them individually [38]. Fig. 9 demonstrates a visual inference test of the trained Vgg19 model on arbitrarily selected six test images to classify the tongue lesions as benign or pre-cancerous. It can be seen that model correctly classified the two pre-cancerous lesions even though both the lesion images are from different perspective. Similarly, the trained ResNET50 model correctly classified the arbitrarily selected six test images into their respective target classes (Fig. 10). Hence we can infer that strongest activations within the Vgg19 and ResNet50 classifiers correspond to the areas of lesions on the tongue. This visual inference technique raises confidence in the ability of the trained DCNN models to successfully identify pre-cancerous tongue lesions.

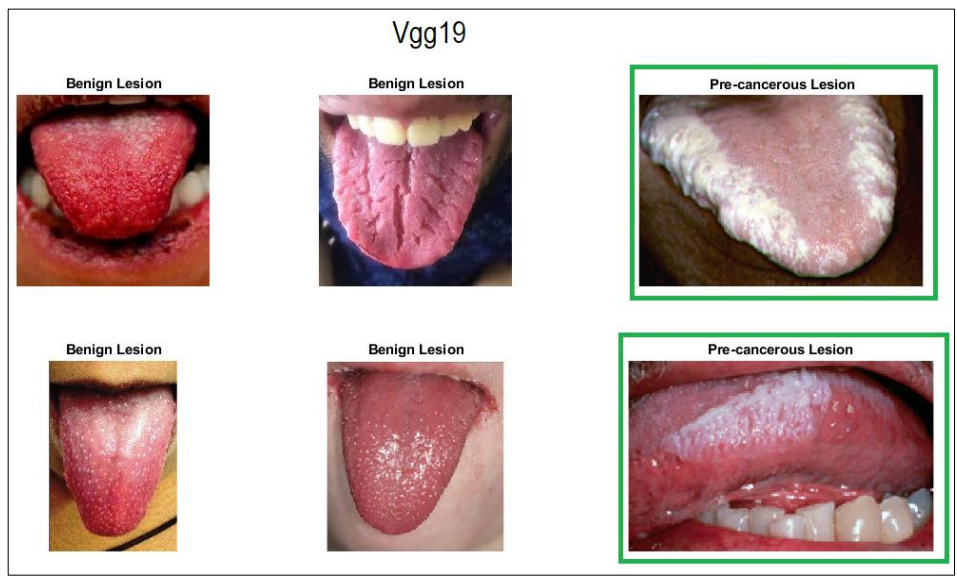

**Fig. 9:** Visual inference test of a trained Vgg19 model on six arbitrarily selected test images for classifying tongue lesions as benign or pre-cancerous.

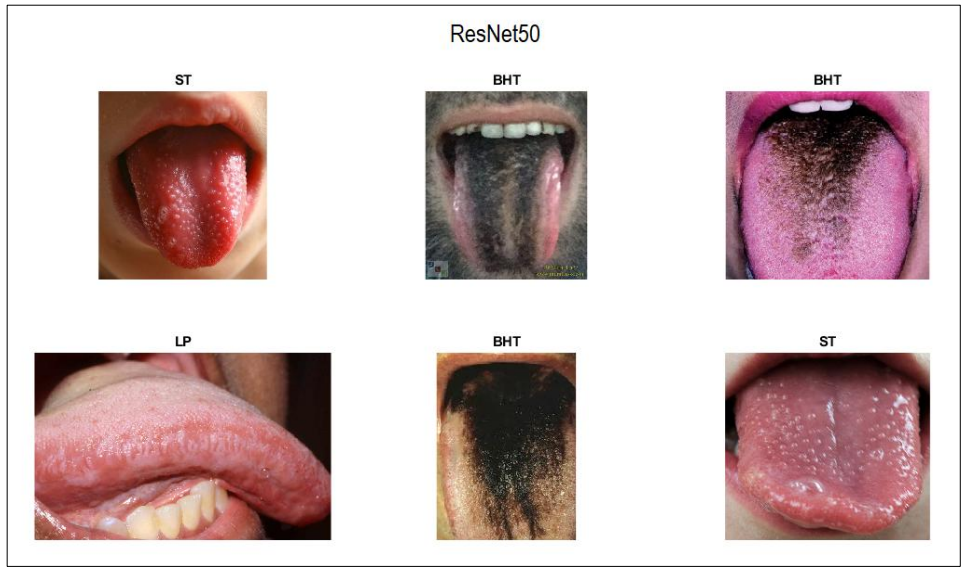

**Fig. 10:** Visual inference test of a trained ResNet50 model on six arbitrarily selected test images for classifying tongue lesions into five target classes.

5. **CONCLUSION**

In this study we have evaluated the efficacy of using DCNNs that can automatically detect pre-cancerous tongue lesions using a small dataset of clinically annotated photographic images, with highest mean binary $A_{CC}$ of 0.98 using the Vgg19 architecture (AUC = 0.9896) and multiclass $A_{CC}$ of 0.97 using the ResNet50 architecture. In both the classification studies, the highest scoring DCNNs had disagreement in 3 of the 80 test cases, which were then blindly reviewed by a certified physician who correctly interpreted all the 3 cases. This (AI + Physician) ensemble approach further enhanced the $A_{CC}$ to 100 % for detecting pre-cancerous tongue lesions. There are limitations to the results presented here. The DCNNs cannot replace human pathological interpretations beyond tongue lesions and in cases where further biopsy investigation is needed. More research is needed to develop the DCNNs to classify lesions occurring in other areas of the oral cavity such as the inner lining of the cheek, due to its close proximity with the tongue. Even though the model presented here is designed for use on tongue lesion cases, it may be applied to lesions occurring in other areas of the oral cavity due to their visual similarity. It is possible that the DCNN trained on tongue lesion image data with the goal of detecting pre-cancerous lesions could flag lesions in other areas of the oral cavity as positive. This work represents a positive step towards an effective and inexpensive OCC pre-screening method that can aid physicians/dentists in their daily clinical practice to triage patients for appropriate clinical management.

**REFERENCES**

1. Bray, F., Ferlay, J. , Soerjomataram, I. , Siegel, R. L., Torre, L. A. and Jemal, A. Global cancer statistics 2018: GLOBOCAN estimates of incidence and mortality worldwide for 36 cancers in 185 countries. *A Cancer Journal for Clinicians* 2018; **68:** 394-424.
2. Oral Cancer. https://www.who.int/cancer/prevention/diagnosis-screening/oral-cancer/en (July 2019).
3. Montero, P. H, and S. G. Patel. Cancer of the oral cavity. *Surgical oncology clinics of North America 2015;* **24:** 491-508.
4. Loo, S., Thunnissen, E., Postmus, P. Wall, I. Granular cell tumor of the oral cavity; a case series including a case of metachronous occurrence in the tongue and the lung. *Medicina oral, patologia oral y cirugia bucal* 2015; **20:** e30-3.
5. Speight, P. M., Khurram, S. A., Kujan, O. Oral potentially malignant disorders: risk of progression to malignancy. *Oral Surgery, Oral Medicine, Oral Pathology and Oral Radiology*, 2018; **125:** 612 – 627.
6. Ho, P. et al. Malignant transformation of oral potentially malignant disorders in males: a retrospective cohort study.  *BMC cancer* 2009; **9:** 260. 10.1186/1471-2407-9-260
7. Foraker, R., Kite, B., Kelley, M., Lai, A. M. EHR-based Visualization Tool: Adoption Rates, Satisfaction, and Patient Outcomes. *eGEMs(Washington, DC)* 2015; **3:** 1159. 10.13063/2327-9214.1159
8. ESTEVA, A., KUPREL, B., NOVOA, R. A., KO, J., SWETTER, S. M., BLAU, H. M., THRUN, S. Dermatologist-level classification of skin cancer with deep neural networks. *Nature* 2017; **542:** 115-118.
9. Hosny, K. M., Kassem M. A., Foaud, M. M. Skin Cancer Classification using Deep Learning and Transfer Learning. *2018 9th Cairo International Biomedical Engineering Conference (CIBEC)*, Cairo, Egypt, 2018; 90-93.
10. Lakhani, P., Sundaram, B. Deep Learning at Chest Radiography: Automated Classification of Pulmonary Tuberculosis by Using Convolutional Neural Networks, *Radiology* 2017; **284:** 574-582.

11. Ma, Ling et al. Deep Learning based Classification for Head and Neck Cancer Detection with Hyperspectral Imaging in an Animal Model, *Proceedings of SPIE--the International Society for Optical Engineering* 2017; **10137:** 101372G; 10.1117/12.2255562.
12. Rajaguru H., Prabhakar S.K. Performance Comparison of Oral Cancer Classification with Gaussian Mixture Measures and Multi Layer Perceptron, *Goh J., Lim C., Leo H. (eds) The 16th International Conference on Biomedical Engineering*, IFMBE Proceedings, Springer Nature, Singapore, 2017; **61:** 123-124.
13. Kim, D. W. et al. Deep learning-based survival prediction of oral cancer patients, *Scientific reports* 2019; **9:** 6994.
14. Lalithmani, K., Punitha, A. Detection of Oral Cancer Using Deep Neural Based Adaptive Fuzzy System in Data Mining Techniques, *International Journal of Recent Technology and Engineering* 2019; **7:** 397-404.
15. Wieslander, H. et al. Deep Convolutional Neural Networks for Detecting Cellular Changes Due to Malignancy, *2017 IEEE International Conference on Computer Vision Workshops (ICCVW)*, Venice, 2017; 82-89.
16. Folmsbee, J., Liu, X., Brandwein-Weber M., Doyle, S. Active deep learning: Improved training efficiency of convolutional neural networks for tissue classification in oral cavity cancer, *2018 IEEE 15th International Symposium on Biomedical Imaging (ISBI 2018)*, Washington, DC, 2018; 770-773.
17. Jeyaraj, P. R., Nadar, E. R. S. Computer-assisted medical image classification for early diagnosis of oral cancer employing deep learning algorithm, *J. Cancer Res. Clin. Oncol.* 2019; **145:** 829; 10.1007/s00432-018-02834-7
18. Aubreville, M. et al. Automatic Classification of Cancerous Tissue in Laserendomicroscopy Images of the Oral Cavity using Deep Learning." *Scientific Reports* **7**, 11979 (2017).
19. Sunil, A. et al. Common Superficial Tongue Lesions, *Indian Journal of Clinical Practice*, 2013; **23:** 534-542.
20. Reamy, B. V. et al. Common Tongue Conditions in Primary Care. *American Family Physician* 2010; **81:** 627-634.
21. Lecun, Y., Bottou, L., Bengio, Y., Haffner, P. Gradient-based learning applied to document recognition, *Proceedings of the IEEE* 1998; **86:** 2278-2324.
22. IMAGENET database, http://www.image-net.org (July 2019)
23. Russakovsky, O., et al. ImageNet Large Scale Visual Recognition Challenge, *Int. J. Comput. Vis.* 2015; **115:** 211.
24. Krizhevsky, A., Sutskever, I., Hinton, G. E. ImageNet classification with deep convolutional neural networks. *Advances in neural information processing systems* 2012; **60:** 1097-1105.
25. C. Szegedy et al. Going deeper with convolutions, *2015 IEEE Conference on Computer Vision and Pattern Recognition (CVPR)*, Boston, MA, 2015; 1-9.

26. Simonyan, K., Zisserman, A. Very Deep Convolutional Networks for Large-Scale Image Recognition, *3rd IAPR Asian Conference on Pattern Recognition (ACPR), Kuala Lumpur* 2015; 730-734.
27. Szegedy, C., Vanhoucke, V., Ioffe, S., Shlens, J., Wojna, Z., Rethinking the Inception Architecture for Computer Vision, *2016 IEEE Conference on Computer Vision and Pattern Recognition (CVPR)*, Las Vegas, NV, 2016; 2818-2826.
28. He, K., Zhang, X., Ren, S., Sun, J. Deep Residual Learning for Image Recognition, *2016 IEEE Conference on Computer Vision and Pattern Recognition (CVPR)*, Las Vegas, NV 2016; 770-778.
29. Iandola, F.N. et al. SqueezeNet: AlexNet-level accuracy with 50x fewer parameters and <0.5MB model size. 2016; Preprint at https://arxiv.org/abs/1602.07360
30. Parikh, R. et al. Understanding and using sensitivity, specificity and predictive values, *Indian journal of ophthalmology* 2008; **56:** 45-50.
31. Wu, R., Yan, S., Shan, Y., Dang, Q., Sun, G. Deep Image: Scaling up Image Recognition, 2016; Preprint at https://arxiv.org/abs/1501.02876
32. Srivastava, N., Hinton, G., Krizhevsky, A., Sutskever, I., Salakhutdinov, R. Dropout: A Simple Way to Prevent Neural Networks from Overfitting, *Journal of Machine Learning Research* 2014; **15:** 1929-1958.
33. Bar, Y., Diamant, I., Wolf, L., Greenspan, H., Deep learning with non-medical training used for chest pathology identification. *SPIE Medical Imaging,* 2015; **94140V:** 10.1117/12.2083124
34. Shin, H., *et al*. Deep Convolutional Neural Networks for Computer-Aided Detection: CNN Architectures, Dataset Characteristics and Transfer Learning, *IEEE Transactions on Medical Imaging*, 2016; **35:** 1285-1298.
35. Hua, K. L., Hsu, C. H., Hidayati, S. C., Cheng, W. H., Chen, Y. J. Computer-aided classification of lung nodules on computed tomography images via deep learning technique. *OncoTargets and therapy* 2015; **8:** 2015–2022.
36. Zhang, W. et al. Deep Convolutional Neural Networks for Multi-Modality Isointense Infant Brain Image Segmentation, *NeuroImage* 2015; **108:** 214-224.
37. Yosinski, J., Clune, J., Nguyen, A., Fuchs, T., Lipson, H., Understanding Neural Networks Through Deep Visualization, Deep Learning Workshop, *31st International Conference on Machine Learning, Lille, France* 2015.
38. Jaeger, S. et al. Two public chest X-ray datasets for computer-aided screening of pulmonary diseases, *Quantitative imaging in medicine and surgery* 2014 **4(6):** 475–477.

**Author Contributions**

M.S.* wrote the main parts of the manuscript. S.S. verified the medical literature in the manuscript. The images utilized in this study were acquired by M. U., M.S. and S.A. and later annotated by S. S.

The analytical experiments were designed by M.S.* and M. S..  M.S.*, S. S. and M. S. analysed the results. S. S. provided medical expertise through discussions. The manuscript was reviewed by all the authors.

**Acknowledgement**

This work was supported by the College of Engineering Scientific Research Center under the Deanship of Scientific Research of King Khalid University.

**Additional Information**

**Competing Interests:** The authors declare that they have no competing interest.